\documentclass[onecolumn, aps, nofootinbib, prd, preprint, floats, floatfix, amsmath, amssymb, superscriptaddress, preprintnumbers]{revtex4-1}
\usepackage{graphicx}
\usepackage{subfigure}
\usepackage{dcolumn}
\usepackage{bm}
\usepackage{amssymb}
\usepackage[utf8]{inputenc}
\usepackage[OT1]{fontenc}
\usepackage{yfonts,amsmath,amsthm,amsfonts,amssymb,amscd, ulem}
\usepackage{enumerate}
\usepackage{fancyhdr}
\usepackage{mathtools}
\usepackage{mathrsfs}
\usepackage{cancel}
\usepackage{slashed}
\usepackage{bigints}
\usepackage[flushleft]{threeparttable}
\usepackage[colorlinks=true, citecolor=purple, linkcolor=blue]{hyperref}
\usepackage{url}
\usepackage{makecell,booktabs}
\usepackage{braket}
\usepackage{relsize}
\usepackage{multirow}
\usepackage{verbatim}
\usepackage{txfonts}
\usepackage{slashed}
\usepackage{upgreek}
\usepackage{extarrows}
\usepackage{array}
\usepackage{appendix}
\usepackage[T1]{fontenc}
\usepackage{setspace}

\usepackage[dvipsnames]{xcolor}

\usepackage{tikz} 
\usetikzlibrary{shapes,arrows,positioning,automata,backgrounds,calc,er,patterns}
\usepackage{tikz-feynman}
\tikzfeynmanset{compat=1.0.0}
\usetikzlibrary{shapes.misc}
\tikzset{cross/.style={cross out, draw=black, fill=none, minimum size=2*(#1-\pgflinewidth), inner sep=0pt, outer sep=0pt}, cross/.default={2pt}}
\usetikzlibrary{shapes.geometric}
\usepackage{orcidlink}

\begin{document}

\title{Probing Purely Inelastic Scalar Dark Matter Across Colliders and Gravitational Wave Observatories}

\author{Jinhui Guo}
\email{guojh23@buaa.edu.cn}
\affiliation{School of Physics, Beihang University, Beijing 100191, China}

\author{Jia Liu \orcidlink{0000-0001-7386-0253}}
\email{jialiu@pku.edu.cn}
\affiliation{School of Physics and State Key Laboratory of Nuclear Physics and Technology, Peking University, Beijing 100871, China}
\affiliation{Center for High Energy Physics, Peking University, Beijing 100871, China}

\author{Chenhao Peng }
\email{chenhaopeng@stu.pku.edu.cn}
\affiliation{School of Physics and State Key Laboratory of Nuclear Physics and Technology, Peking University, Beijing 100871, China}

\author{Xiao-Ping Wang \orcidlink{0000-0002-2258-7741} }
\email{hcwangxiaoping@buaa.edu.cn}
\affiliation{School of Physics, Beihang University, Beijing 100083, China}

\preprint{CPTNP-2025-026}

\begin{abstract}
We propose and study a purely inelastic scalar dark matter model, where two real scalars-dark matter $\phi_1$ and its excited partner $\phi_2$ interact with the Standard Model via a Higgs portal. After mass diagonalization, only inelastic couplings remain, allowing the model to evade stringent bounds from direct detection. We show that thermal (co-)annihilation between $\phi_1$ and $\phi_2$ naturally yields the observed dark matter relic abundance. The same interaction structure can induce a strongly first-order phase transition in the early universe, generating detectable gravitational waves in upcoming experiments. Meanwhile, the slight mass splitting between $\phi_1$ and $\phi_2$, along with the heavy off-shell mediator SM Higgs, leads to long-lived particle signatures of $\phi_2$ at the HL-LHC via the displaced muon-jets technique. 
We pinpoint a feasible parameter space where the correct relic abundance, observable gravitational waves, and collider signals can all be achieved concurrently, presenting a valuable chance to validate this scenario through a comprehensive examination encompassing cosmological, astrophysical, and collider investigations.
\end{abstract}

\maketitle

\tableofcontents

\section{Introduction}
\label{sec:int}
Following the discovery of the Higgs boson at the LHC, understanding the properties of the electroweak phase transition (EWPT) has emerged as a significant challenge in modern particle physics \cite{ATLAS:2012yve,CMS:2012qbp,Athron:2023xlk}. Lattice studies within the framework of the Standard Model (SM) indicate that the EWPT proceeds as a continuous crossover rather than a first-order phase transition \cite{Kajantie:1996qd,Rummukainen:1998as,Laine:1998jb}. However, in various theoretical frameworks beyond the SM, such as the real singlet scalar extension to the SM (xSM)~\cite{Profumo:2007wc,Espinosa:2011ax,Cline:2012hg,Alanne:2014bra,Profumo:2014opa,Alves:2018jsw,Vaskonen:2016yiu,Huang:2018aja,Cheng:2018ajh,Alanne:2019bsm,Gould:2019qek,Carena:2019une,Ghorbani:2018yfr,Ghorbani:2017jls,Goncalves:2024vkj,Chatterjee:2022pxf,Ghosh:2022fzp}, the two-Higgs-doublet model~\cite{Dorsch:2013wja,Chao:2015uoa,Basler:2016obg,Haarr:2016qzq,Dorsch:2017nza,Andersen:2017ika,Bernon:2017jgv,Wang:2018hnw,Wang:2019pet,Kainulainen:2019kyp,Su:2020pjw,Bittar:2025lcr,DiazSaez:2024nrq,Goncalves:2023svb}, and others, the inclusion of additional ``scalars'' alongside the SM Higgs field can lead to a first-order electroweak phase transition (FOEWPT). This FOEWPT has the potential to push the early Universe out of thermal equilibrium, creating conditions conducive to electroweak baryogenesis \cite{Morrissey:2012db,Cline:2006ts,Trodden:1998ym}, a process explaining the observed matter-antimatter asymmetry in the universe. Furthermore, a first-order electroweak phase transition may generate detectable gravitational waves (GWs), offering prospects for future GW observatories.

The newly introduced scalar can potentially act as a dark matter (DM) candidate. Despite the remarkable success of the SM in explaining various phenomena observed in particle experiments and astrophysical studies, the mystery surrounding the nature of dark matter remains a significant unresolved issue \cite{Roszkowski:2017nbc,Planck:2018vyg,ParticleDataGroup:2014cgo}. Among the different candidates, the Weakly Interacting Massive Particle (WIMP) framework is notable for its ability to naturally account for the observed dark matter relic abundance, with $\Omega h^2 = 0.1198 \pm 0.0026$ \cite{Planck:2018vyg}, achieved through thermal freeze-out with an annihilation cross-section typical of the electroweak scale. This intriguing correlation hints at the possibility of new physics emerging at or above the weak scale. The validity of the WIMP paradigm has been extensively assessed using various methods, including direct detection \cite{XENON:2018voc,CDEX:2019hzn,LUX:2016ggv,PandaX-4T:2021bab,PandaX-II:2017hlx,LZ:2022lsv}, indirect detection \cite{AMS:2014xys,Fermi-LAT:2013sme,DAMPE:2017fbg,AMS:2014bun,Fermi-LAT:2015att,Fermi-LAT:2016uux,DAMPE:2017fbg,Super-Kamiokande:2004pou,Super-Kamiokande:2011wjy}, and collider searches \cite{Abdallah:2015ter,Abercrombie:2015wmb,Krnjaic:2015mbs,Boveia:2018yeb,Arcadi:2019lka,ATLAS:2025auy,Pokidova:2025jvq,Wassmer:2025kxt}. Nevertheless, the absence of definitive signals has led to a growing interest in exploring alternative dark matter scenarios beyond the conventional WIMP framework.

An appealing alternative to the traditional WIMP scenario is the coannihilation mechanism \cite{Griest:1990kh, Baker:2015qna}, which introduces an additional, heavier state named the coannihilation partner or dark matter partner that is closely related to the dark matter particle. In the specific case of an inelastic dark matter (iDM) model, the coannihilation partner corresponds to an excited state, while the stable dark matter resides in the ground state \cite{Tucker-Smith:2001myb, Tucker-Smith:2004mxa,DallaValleGarcia:2024zva,Voronchikhin:2025eqm,Zhang:2024sox,DallaValleGarcia:2024zva,Berlin:2023qco,Lu:2023cet,Izaguirre:2015zva,DEramo:2016gqz,Berlin:2018jbm,Okada:2019sbb,Kang:2021oes,Bell:2021zkr,Batell:2021ooj,Bell:2021xff,Li:2021rzt,Filimonova:2022pkj,Bertuzzo:2022ozu}.
These partners have the capability to interact with SM particles and, when combined with dark matter, can collectively undergo annihilation to produce SM particles. In this configuration, the strength of interaction between dark matter and SM particles can be significantly reduced or even nonexistent, leading to a diminished annihilation cross section for pairs of dark matter. This characteristic inherently aids in steering the model clear of stringent constraints imposed by observations from the Cosmic Microwave Background (CMB) \cite{Planck:2018vyg,Slatyer:2015jla} and various indirect detection experiments~\cite{AMS:2014xys,AMS:2014bun,Fermi-LAT:2015att,Fermi-LAT:2016uux,DAMPE:2017fbg}, which are sensitive to energy injection from late-time annihilations. Moreover, the feeble coupling to SM particles also enables the model to effectively bypass the limitations set by direct detection conducted in deep underground facilities \cite{LUX:2016ggv,CDEX:2019hzn,XENON:2018voc,PandaX-4T:2021bab}.

In this research, we investigate a model of ``purely'' inelastic dark matter interaction involving the coannihilation mechanism. The dark sector comprises a complex scalar denoted as $\phi=\frac{1}{\sqrt{2}}(\phi_1+i\phi_2)$, which interacts with SM particles through the Higgs boson. Here 'purely inelastic' means that only inelastic scattering survives, i.e. all tree-level $\phi_i-\phi_i-H$ couplings vanish.
Previous studies on inelastic dark matter have mainly focused on light scalar or fermionic models with masses below 100 GeV. These works explored signals at present and future colliders, neutrino detectors, or astrophysical observations, and emphasized how the models account for the observed dark matter relic abundance \cite{Izaguirre:2015zva,Berlin:2018jbm,Batell:2021ooj,Li:2021rzt,Bertuzzo:2022ozu,Lu:2023cet,Garcia:2024uwf,Krnjaic:2024ols,Okada:2024kni,Zhang:2024sox,Liu:2025abt}. Other works studied heavier iDM models above 100 GeV, where annihilation between coannihilation partners dominates \cite{Guo:2021vpb,Hooper:2025fda}. These models can be probed at the upcoming High-Luminosity LHC (HL-LHC), and typically require moderate couplings between the dark sector and the Standard Model Higgs. Some studies also considered fermionic iDM models with additional scalar mediators, which can induce strongly first-order phase transitions and detectable collider signatures \cite{Mirzaie:2025bzn}.
In contrast, our work, with a straightforward purely iDM model, not only addresses the dark matter relic abundance and (in-)direct detection signals, but also naturally includes possible first-order phase transitions and the consequent gravitational wave signals arising from the newly introduced scalar. Furthermore, if the mass splitting between the dark matter particle $\phi_1$ and its excited partner $\phi_2$ is small, the latter can behave as a long-lived particle (LLP) at collider scales \cite{Alimena:2019zri}. In this case, hadron colliders can search for LLP signatures using displaced muon-jets or time-delayed techniques, which provide powerful tools for suppressing QCD backgrounds.

The structure of this paper is outlined as follows. In Sec. \ref{sec:model}, the scalar iDM model is discussed, and decay channels and lifetime of the DM partner are also calculated. In Sec. \ref{sec:const}, we investigate the DM relic abundance and various existing constraints from (in-)direct detection, thermalization requirement, collider searches, and Higgs precision measurements. In Sec. \ref{sec:FOPT}, we explore the electroweak phase transition with the new scalars and their possible gravitational waves. In Sec. \ref{sec:collider}, the long-lived signatures of the DM partner $\phi_2$ are studied. In Sec. \ref{sec:concl}, we conclude.

\section{The model}\label{sec:model}
In this work, we investigate a model of inelastic DM, which involves a complex scalar field, $\hat\phi$. This field is composed of two real scalar fields: $\hat\phi_1$ (the ground state) and $\hat\phi_2$ (the excited state). We place a hat on the field to distinguish it from its mass eigenstates.
The excited state, $\hat\phi_2$, serves as a DM partner of $\hat\phi_1$. The complex scalar field $\phi$ interacts with the SM Higgs field, $H$, via a quadratic coupling, and this coupling is in general complex. The interaction is described by the following Lagrangian:
\begin{equation}
\begin{aligned}
\mathcal{L} =& \left(\partial_\mu \hat\phi\right)^\dagger \left(\partial^\mu\hat\phi\right)-\frac{1}{2}\mu_1^2\hat\phi_1^2-\frac{1}{2}\mu_2^2\hat\phi_2^2-\lambda_\phi\left(\hat\phi^\dagger\hat\phi\right)^2-2\lambda_I\hat\phi^2|H|^2+h.c.\\  
& +\mu^2 H^\dagger H-\lambda\left(H^\dagger H\right)^2,
\label{eq:Lagrangian-1}
\end{aligned}
\end{equation}
where $\lambda_I = \lambda_{I,r} + i\lambda_{I,i}$ is the complex coupling between the scalar field $\phi$ and the Higgs field, with $\lambda_{I,r}$ and $\lambda_{I,i}$ representing the real and imaginary components, respectively. Through this Higgs portal interaction, DM and its coannihilation partner interact with SM particles, mediated by the Higgs field. The Higgs field and the complex scalar $\hat\phi$ are expressed after electroweak symmetry breaking as:
\begin{align}
H&=\frac{1}{\sqrt{2}}
\begin{pmatrix}H^+\\ v_h+h+i\chi\end{pmatrix}, ~~~\hat\phi  = \frac{1}{\sqrt{2}}\left(\hat\phi_1+i\hat\phi_2 \right).
\end{align}
In this setup, only the Higgs field develops a vacuum expectation value (VEV), $v_h$, whereas the complex scalar $\hat\phi$ does not, protected by a discrete $Z_2$ symmetry under which only $\hat\phi$ is odd.
The Goldstone bosons, $H^\pm$ and $\chi$, are absorbed by the SM gauge bosons. After electroweak symmetry breaking, the mass matrix for $\hat\phi_1$ and $\hat\phi_2$ can be written as:
\begin{equation}\label{eq:mMatrix}
M = 
\begin{pmatrix}
\mu_1^2+2\lambda_{I,r}v_h^2&-2\lambda_{I,i}v_h^2\\
-2\lambda_{I,i}v_h^2&\mu_2^2-2\lambda_{I,r}v_h^2
\end{pmatrix}.
\end{equation}
This mass matrix $M$ can be diagonalized through a unitary rotation, $U$, such that:
\begin{equation}
~U M U^\dagger={\rm diag}(m_1^2,m_2^2), 
\end{equation}
where the rotation matrix $U$ is given by:
\begin{equation} U=\begin{pmatrix}\cos\theta&-\sin\theta\\\sin\theta&\cos\theta\end{pmatrix},
\end{equation}
with the rotation angle satisfying
\begin{equation}\label{eq:tantheta1}
\tan\left(2\theta\right)= \frac{4\lambda_{I,i}v_h^2}{4\lambda_{I,r}v_h^2+\mu_1^2-\mu_2^2}.
\end{equation}
With this rotation, we obtain the mass eigenstates $\phi_1$ and $\phi_2$ from the flavor eigenstates $\hat\phi_1$ and $\hat\phi_2$ as follows:
 \begin{equation}
 ~\begin{pmatrix}
\phi_1\\ \phi_2
\end{pmatrix}=U\cdot \begin{pmatrix}
\hat\phi_1\\ \hat\phi_2
\end{pmatrix}.
 \end{equation}
After rotating from the flavor eigenstates to the mass eigenstates, the couplings among mass eigenstates also need to be modified, which is relevant to the interaction term
\begin{align}
\Delta \mathcal L&=-
\begin{pmatrix}
\hat\phi_1&\hat\phi_2    
\end{pmatrix}
\begin{pmatrix}
\lambda_{I,r}&-\lambda_{I,i}\\
-\lambda_{I,i}&-\lambda_{I,r}
\end{pmatrix}
\begin{pmatrix}
\hat\phi_1\\ \hat\phi_2
\end{pmatrix}
\left(2 v_h h + h^2 \right)=\begin{pmatrix}
\hat\phi_1&\hat\phi_2    
\end{pmatrix}
\hat \Lambda
\begin{pmatrix}
\hat\phi_1\\ \hat\phi_2
\end{pmatrix}
\left(2 v_h h + h^2 \right)\\
&=\begin{pmatrix}
\phi_1&\phi_2    
\end{pmatrix}
U\hat\Lambda U^\dagger
\begin{pmatrix}
\phi_1\\ \phi_2
\end{pmatrix}
\left(2 v_h h + h^2 \right)
\end{align}
where
\label{eq:coupling-phys}
\begin{align}
\Lambda =&U\hat\Lambda U^\dagger =
\begin{pmatrix}
\lambda_{11} & \lambda_{12}\\
\lambda_{21} & \lambda_{22}\\
\end{pmatrix}=\cos(2\theta)\cdot\begin{pmatrix}
 \lambda_{I,i}  \tan (2\theta ) + \lambda_{I,r}  & \tan (2\theta)  \lambda_{I,r} -  \lambda_{I,i} \\
 \tan (2\theta)   \lambda_{I,r}  -  \lambda_{I,i} & - \lambda_{I,i}  \tan (2\theta) - \lambda_{I,r}\\
\end{pmatrix}
\end{align}
It is clear that the coupling matrices, $\Lambda$ and $\hat \Lambda$, are both traceless, which implies that the interactions between $\hat\phi_1 \hat\phi_1 h$ and $\hat\phi_2 \hat\phi_2 h$ have opposite signs, both before and after diagonalization. 

Direct-detection (DD) experiments impose stringent upper limits on DM interactions with nucleons. In our framework, the quartic coupling $\lambda_{11}$ induces a Higgs-mediated spin-independent (SI) scattering cross section,
\begin{equation}
\sigma_{\phi_1 N}^{\mathrm{SI}} = \frac{4\lambda_{11}^2}{\pi m_h^4}\frac{m_N^4 f_N^2}{(m_1+m_N)^2},
\end{equation}
where $m_N$ denotes the nucleon mass and $f_N\simeq0.3$. The most recent XENONnT and LZ results \cite{XENON:2025vwd,LZ:2024zvo} place the strongest constraints in the mass range relevant to our study, requiring $\sigma_{\phi_1 N}^{\mathrm{SI}}\lesssim10^{-48}~{\rm cm}^2$ around $m_1\simeq40$ GeV. Translating these bounds into limits on the quartic coupling yields $\lambda_{11}\lesssim3.7\times10^{-4}$ for $m_1\simeq10$ GeV, $\lambda_{11}\lesssim10^{-4}$ for $m_1\simeq40$ GeV, and $\lambda_{11}\lesssim1.2\times10^{-4}$ for $m_1\simeq60$ GeV \cite{Arcadi:2019lka}. To explicitly satisfy these constraints and simplify the analysis, we take $\lambda_{11}=0$, which is a natural and convenient choice within our framework. Thus, we require that the transformation angle $\theta$ satisfy the condition:
\begin{equation}
\tan\left(2\theta\right) = -\frac{\lambda_{I,r}}{\lambda_{I,i}}.\label{eq:tantheta2}
\end{equation}
Substituting this into Eq. (\ref{eq:tantheta1}), we obtain the following relation:
\begin{align}
-\frac{\lambda_{I,r}}{\lambda_{I,i}}
=\frac{4\lambda_{I,i}v_h^2}{4\lambda_{I,r}v_h^2+\mu_1^2-\mu_2^2},
\end{align}
which leads to 
\begin{align}
\lambda_{I,i}=\pm \frac{1}{2v_h}\sqrt{(\mu_2^2-\mu_1^2-4\lambda_{I,r}v_h^2) \lambda_{I,r}}.
\end{align}
With this relationship, the coupling matrix will be simplified to:

\begin{align}
\Lambda 
=\begin{pmatrix}
0& \mp\frac{1}{2v_h}\sqrt{\left(\mu_2^2-\mu_1^2\right) \lambda_{I,r} }\\
\mp\frac{1}{2v_h}\sqrt{\left(\mu_2^2-\mu_1^2\right) \lambda_{I,r} } & 0
\end{pmatrix},
\end{align}
and the physical masses of $\phi_1$ and $\phi_2$ are
\begin{align}
m_1^2 &=\frac{1}{2}\Bigl(\mu_{1}^{2}+\mu_{2}^{2}
-\sqrt{\bigl(\mu_{1}^{2}-\mu_{2}^{2}+4\,\lambda_{\mathrm{I,r}}\,v_{h}^{2}\bigr)\,
\Bigl(\mu_{1}^{2}-\mu_{2}^{2}\Bigr)}\;\Bigr) ,\\
m_2^2 &=\frac{1}{2}\Bigl(\mu_{1}^{2}+\mu_{2}^{2}
+\sqrt{\bigl(\mu_{1}^{2}-\mu_{2}^{2}+4\,\lambda_{\mathrm{I,r}}\,v_{h}^{2}\bigr)\,
\Bigl(\mu_{1}^{2}-\mu_{2}^{2}\Bigr)}\;\Bigr).
\end{align}
\indent By nullifying the direct interaction coupling $\lambda_{11}$, we get a purely iDM model. In the mass eigenstate basis, the effective Lagrangian can be written as:
\begin{equation}\label{eq:physical-lagrangian}
\begin{aligned}
\mathcal{L} =&
\frac{1}{2}\partial_\mu\phi_1\partial^\mu \phi_1
+\frac{1}{2}\partial_\mu\phi_2\partial^\mu \phi_2
-\frac{1}{2}m_1^2\phi_1^2 - \frac{1}{2}m_2^2 \phi_2^2 - \lambda_{12}\phi_1\phi_2(2v_h h+ h^2)-
\lambda_\phi (\phi^\dagger\phi)^2 - V(h),
\end{aligned}
\end{equation}
where
\begin{equation}
\begin{aligned}
    \lambda_{12} &=\mp\frac{\sqrt{(\mu_2^2-\mu_1^2)\lambda_{I,r}}}{v_h}.
\end{aligned}
\end{equation}
We can introduce a dimensionless parameter to represent the mass splitting between the two real scalar fields:
\begin{equation}
\Delta \equiv \frac{m_2-m_1}{m_1},
\end{equation}
Thus, there are 4 free physical parameters of Lagrangian (\refeq{eq:physical-lagrangian}):
\begin{equation}
    \{m_1,\Delta,\lambda_{12},\lambda_{\phi} \}. \label{eq:parameters}
\end{equation}
\indent In the rest of the paper, we occasionally retain the notation $m_2$ to simplify the mathematical expressions.
Finally, we can express the initial parameters in terms of the physical ones using the following relations: 
\begin{align}
\lambda_{I,r} &= \frac{v_h^2\lambda_{12}^2}{\sqrt{(m_2^2-m_1^2)^2+4v_h^4\lambda_{12}^2}} ,\\
\lambda_{I,i} &= \pm\frac{1}{2v_h}\sqrt{(\mu_2^2-\mu_1^2-4\lambda_{I,r}v_h^2)\lambda_{I,r}} ,
\end{align}
\begin{align}
\mu_2^2 &= \frac{1}{2}\left((m_1^2+m_2^2)+\sqrt{(m_2^2-m_1^2)^2+4v_h^4\lambda_{12}^2}\right),\\
\mu_1^2 &= \frac{1}{2}\left((m_1^2+m_2^2)-\sqrt{(m_2^2-m_1^2)^2+4v_h^4\lambda_{12}^2}\right).
\label{eq:parameter-relations}
\end{align}

\subsection{Decay Processes and Branching Ratios of Inelastic Dark Matter Partners}
Due to the interaction between $\phi_1$, $\phi_2$ and Higgs, $\phi_2$ can decay into $\phi_1  h$ if kinematically allowed. In our study, we are interested in $\Delta \times m_1< m_h$, so that $\phi_2$ can only decay into an off-shell Higgs, which then undergoes a two-body decay into SM particles. 

The decay width of $\phi_2$ is influenced by several factors, including the small mass difference $\Delta$, the weak Yukawa coupling between SM fermions and the Higgs boson, and the presence of an off-shell Higgs mass. These factors may result in $\phi_2$ being long-lived under certain conditions. In the following analysis, we will systematically calculate the decay width of $\phi_2$ for each decay channel. The decay process can generally be categorized into two main channels: the leptonic and the hadronic channels.

Before delving into the calculations, it is crucial to consider the scenario where the mass difference is significantly larger than the relevant energy scale (2 GeV) of the perturbative spectator model \cite{Grinstein:1988yu,Winkler:2018qyg} and the fermion mass, $\Delta\cdot m_1\gg 2$ GeV and $\Delta\cdot m_1\gg 2m_{f}$, where the partial decay width of $\phi_2\to\phi_1 f\bar{f}$ can be reasonably approximated as follows:
\begin{equation}
    \Gamma(\phi_2\to\phi_1 f\bar{f})\simeq \frac{\lambda_{12}^2m_f^2m_2^3\Delta^5}{60\pi^3m_h^4}\times\theta(m_1\cdot\Delta-2m_f),
\end{equation}
where the fermion mass $m_f$ is neglected for phase space integration. This approximation remains valid for the leptonic decay channels as long as $m_1\cdot \Delta\gg2m_f$, even when $m_1\cdot \Delta<2$ GeV. 

However, when the fermion mass $m_f$ is comparable to the mass splitting $m_1\cdot \Delta$, numerical integration over the phase space is necessary. 
For the kinematically allowed leptonic decay channel, the complete decay width can be determined as:
\begin{equation}
    \Gamma(\phi_2\to\phi_1 \ell^+\ell^-)=\frac{8m_\ell^2 \lambda_{12}^2}{(2\pi)^3 32m_2^3}\int_{4m_\ell^2}^{m_1^2\Delta^2}\frac{(s_{12}-4m_\ell^2)^{3/2}\sqrt{\frac{m_1^4+(m_2^2-s_{12})^2-2m_1^2(m_2^2+s_{12})}{s_{12}}}}{(s_{12}-m_h^2)^2} ds_{12},
\end{equation}
where $s_{12}$ represents the invariant mass of the final lepton pair. Specifically, we focus on the muon final state due to its considerable branching ratio and distinct signals observable at the HL-LHC. And the SM backgrounds can be effectively reduced through targeted selection criteria.

For mass differences smaller than 2 GeV, the calculation becomes somewhat intricate and requires careful consideration, especially concerning the types of fermions present. In the case of hadronic decay channels, only the total hadronic decay width is of significance, as it allows for the determination of the overall decay width. These calculations often require accounting for non-perturbative effects.

Considering Ref. \cite{Winkler:2018qyg}, the integral range of $s_{12}$ can be split into two segments: one from $(2m_{\rm final})^2$ to $(2~\rm GeV)^2$, where $m_{\rm final}$ is the mass of the SM particle from $\phi_2$ decays and $2m_{\rm final}<2$ GeV, and the other extending beyond $(2~\rm GeV)^2$. In the former range, the methodology of chiral perturbation theory elucidated in Ref. \cite{Winkler:2018qyg} is utilized to compute the amplitudes for diverse decay processes, such as $\phi_2\to\phi_1\pi\pi$, $\phi_2\to\phi_1 KK$, $\phi_2\to\phi_1 \eta\eta~(\rho\rho)$, among others. In the latter range, the parton-level amplitude is good enough, which can be directly derived following the Feynman rules of QCD. In accordance with Ref. \cite{Winkler:2018qyg}, all feasible channels, like $\phi_2\to\phi_1 gg$, $\phi_2\to\phi_1 c\bar{c}$, $\phi_2\to\phi_1 s\bar{s}$, $\phi_2\to\phi_1 b\bar{b}$, are computed. By summing the contributions from all relevant decay channels in each interval and integrating over $s_{12}$ for each segment separately, the total width of the hadronic channel can be determined.

One might wonder whether the photon channel should be included in the calculation of the total decay width, similar to the significance of the SM Higgs discovery at the LHC \cite{ATLAS:2012yve}, where the branching ratio $Br(h\to\gamma\gamma)$ is around $0.2\%$. In our model, it is observed that the branching ratio of $\phi_2\to\phi_1\gamma\gamma$ falls within the range of $[2.5\times10^{-5},2.6\times10^{-4}]$ for $m_2\in[50,120]$ with $\Delta=0.05$, significantly smaller than the decay width of channel $\phi_2\to\phi_1 gg$ by about three orders of magnitude. While the photon channel benefits from a clean SM background, it suffers from a much lower branching ratio and greater challenges in track reconstruction compared to the muon channel. For this reason, our collider study will focus on the muon channel, and the determination of the total decay width and branching ratio for the muon pair will take priority.\\
\begin{figure}[htbp]
    \centering
    \includegraphics[width=0.48\textwidth]{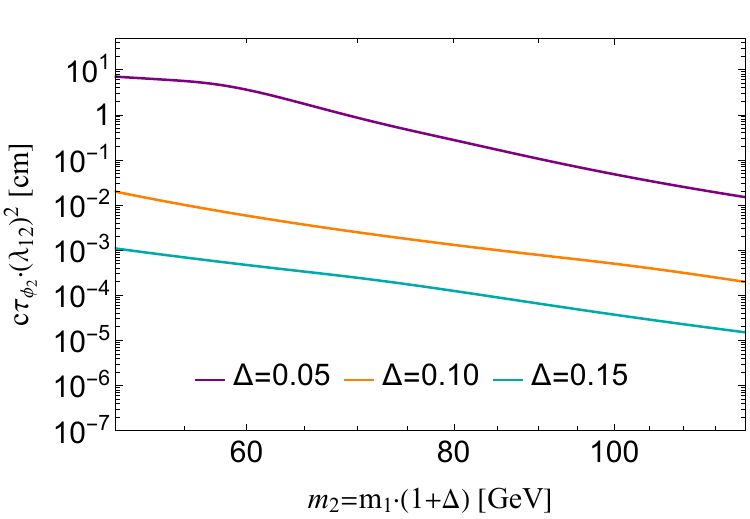}
    \includegraphics[width=0.48\textwidth]{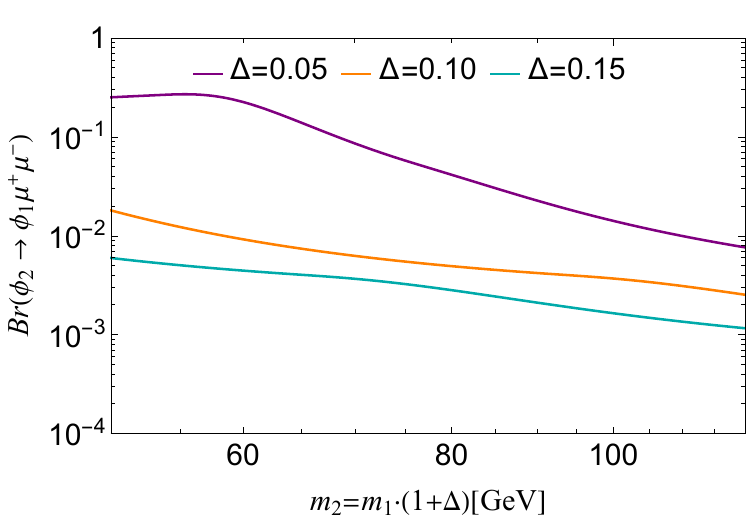}
    \caption{Total decay width and branching ratio of muon pair final state of DM partner as a function of its mass for three different mass splittings. The purple line represents the results of mass splitting $\Delta=0.05$, the orange one represents the results of $\Delta=0.10$, and the cyan one corresponds to $\Delta=0.15$.}
    \label{fig:gamma-br}
\end{figure}
\indent In Fig. \ref{fig:gamma-br}, we show the proper decay length (left panel), $c\tau_{\phi_2}=1/\Gamma_{tot}$, and the branching ratio 
$Br(\phi_2\to\phi_1\mu^+\mu^-)$ (right panel) for three different mass splittings. For the mass range of interest, $m_{2}\in[50,110]$ GeV, the rescaled proper decay length for $\Delta=0.05$ spans from $10^{-2}$ cm to 7 cm. This suggests that the dark matter partner $\phi_2$ could be long-lived at collider scales, provided that $\lambda_{12}$ is on the order of 0.1. However, for larger values of $\Delta$, the decay length decreases by several orders of magnitude, making it challenging to observe a displaced or delayed signal unless $\lambda_{12}$ is extremely small. Regarding the branching ratio $Br(\phi_2\to\phi_1\mu^+\mu^-)$, it stretches from 1\% to 30\% for the cases shown. In contrast, for the other two scenarios, the branching ratios are significantly smaller, making detection more difficult at colliders. Another concern is whether prompt decays of the DM partners could be observed, especially in scenarios with large mass differences. In particular, prompt decays leading to muon pairs could produce signals like monojet plus a muon pair, accompanied by substantial missing transverse energy, $j+\slashed{E}_T+\mu^+\mu^-$, is feasible. However, these prompt decay signatures are often overwhelmed by large Standard Model backgrounds, complicating detection at proton-proton colliders. Therefore, we focus initially on the long-lived signatures of the DM partner, with the exploration of prompt decay signatures left for future research. 

It is important to note that the branching ratios of alternative decay channels, such as $\phi_2 \rightarrow \phi_1 \tau^{+} \tau^{-}$, $\phi_2 \rightarrow \phi_1 \pi \pi$, and other jet channels, can also be computed. Particularly significant are the $\phi_2 \rightarrow \phi_1 c \bar{c}$ and $\phi_2 \rightarrow \phi_1 \tau^{+} \tau^{-}$channels, which become dominant for $m_2 \gtrsim 60 \mathrm{GeV}$ and 80 GeV , respectively. However, due to the small mass differences involved, the jets produced in these decays are too soft to be detected, and therefore, these channels are not considered further here. In contrast, the muon final state benefits from a lower $p_T$ trigger, which facilitates easier reconstruction and better separation from QCD backgrounds. This makes the muon channel a highly effective search channel at $p p$ colliders. A more detailed exploration of collider searches will be provided in the following section.

\section{Dark Matter and Existing Constraints}
\label{sec:const}
\subsection{Dark Matter Relic Abundance}
In our model, both $\phi_1$ and $\phi_2$ can contribute to the DM relic abundance via the coannihilation process~\cite{PhysRevD.43.3191}. Assuming that the number density ratio between $\phi_1$ and $\phi_2$ follows its equilibrium value, the system reduces to an effective single-species Boltzmann equation, which can be solved using the effective annihilation cross section,
\begin{equation}
\sigma_{\rm eff}=\frac{g_{\phi_1}^2}{g_{\rm eff}^2}\left(\sigma_{11}+2\sigma_{12}\frac{g_{\phi_2}}{g_{\phi_1}}\left(1+\Delta\right)^{3/2}e^{-x\cdot\Delta}+\sigma_{22}\frac{g_{\phi_2}^2}{g_{\phi_1}^2}\left(1+\Delta\right)^3e^{-2x\cdot \Delta} \right),
\end{equation}
where $\sigma_{\ij} = \sigma(\phi_i \phi_j \to \mathrm{SM~SM})$ denotes the annihilation cross section into SM particles, $g_{\phi_1} = g_{\phi_2}$ represent the intrinsic degrees of freedom of $\phi_1$ and $\phi_2$, $x = m_1 / T$ with $T$ being the temperature of the thermal bath. The effective degree of freedom $g_{\rm eff}$ is defined as
\begin{equation}
g_{\rm eff} = g_{\phi_1}+g_{\phi_2}\left(1+\Delta\right)^{3/2}e^{-x\cdot \Delta}.
\end{equation}
\begin{figure}[htbp]
    \centering
    \includegraphics[width=1.0\linewidth,trim=0 100 40 100,clip]{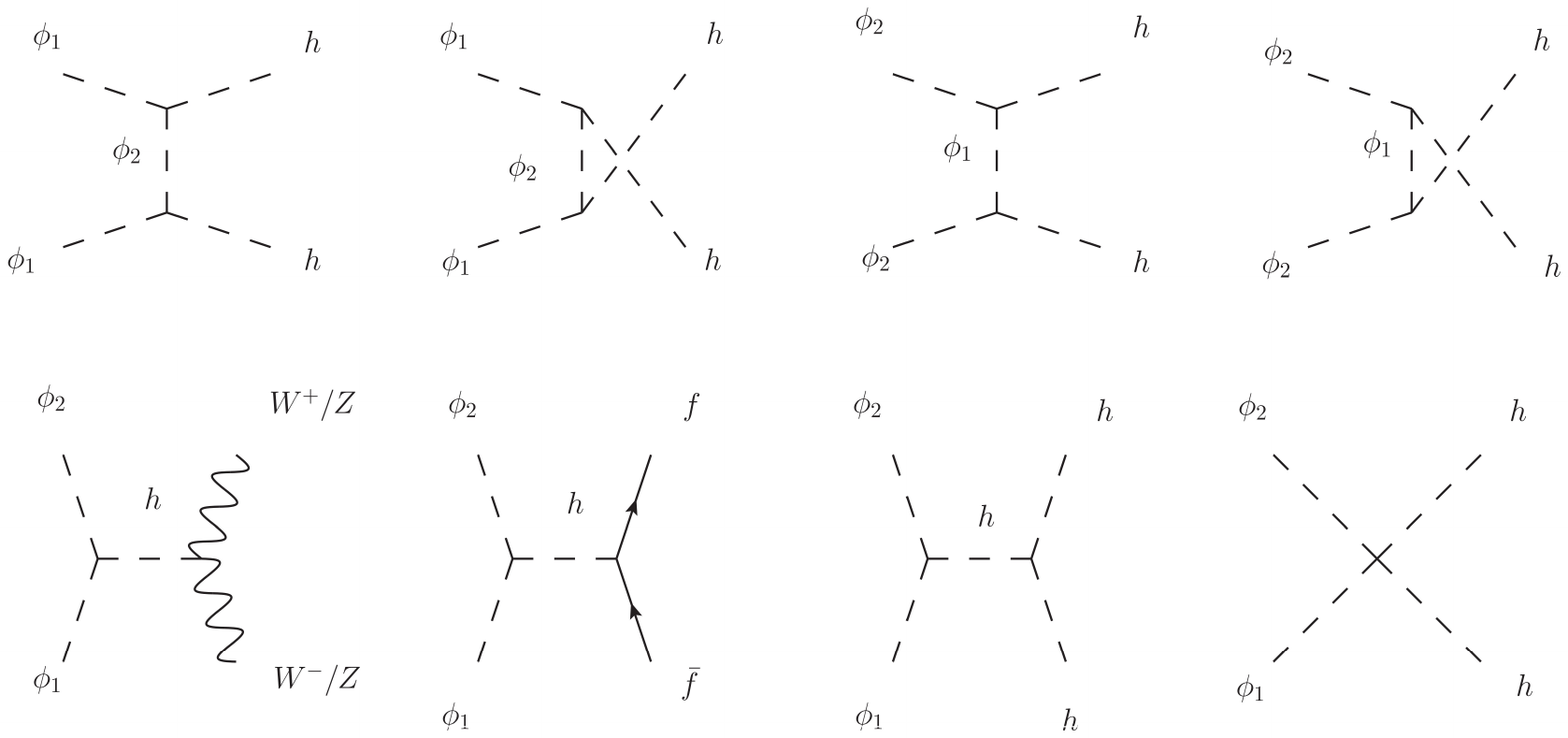}
    \caption{Feynman diagrams for the annihilation processes $\phi_i\phi_j\to {\rm SM~SM}$}
    \label{fig:annihilation-Feynman-diagram}
\end{figure}

The annihilation processes $\phi_i \phi_j \to \textrm{SM}~\textrm{SM}$ include final states such as $\bar{f}f$, $W^+W^-$, $ZZ$, and $hh$. Their corresponding Feynman diagrams are shown in Fig. \ref{fig:annihilation-Feynman-diagram}. The total $s$-wave contributions to the annihilation cross sections, denoted by $\langle \sigma v \rangle_s$, can be computed as follows:
\begin{equation}
\langle\sigma v\rangle_{s} =  \langle\sigma v\rangle_{f\bar{f}}+\langle\sigma v\rangle_{WW}+\langle\sigma v\rangle_{ZZ}+\langle\sigma v\rangle_{hh},
\end{equation}
which includes four processes are proportional to $\lambda_{12}^2$
\begin{equation}
\begin{aligned}
\langle \sigma v\rangle_{f\bar{f}} &\simeq \frac{\lambda_{12}^2 \, m_f^2 \left((m_1 + m_2)^2 - 4\, m_f^2\right)^{3/2}}{2\pi \sqrt{m_1 \, m_2} \, (m_1 + m_2)^2 \, (m_1 + m_2 - m_h)^2 \, (m_1 + m_2 + m_h)^2},\\
\langle\sigma v\rangle_{WW}&\simeq
\frac{\lambda_{12}^2 \sqrt{(m_1 + m_2)^2 - 4 m_W^2} \left(-4 m_W^2 (m_1 + m_2)^2 + (m_1 + m_2)^4 + 12 m_W^4\right)}{4\pi \sqrt{m_1 m_2} (m_1 + m_2)^2 (m_1 + m_2 - m_h)^2 (m_1 + m_2 + m_h)^2},\\
\langle\sigma v\rangle_{ZZ}&\simeq
\frac{\lambda_{12}^2 \sqrt{(m_1 + m_2)^2 - 4 m_Z^2} \left(-4 m_Z^2 (m_1 + m_2)^2 + (m_1 + m_2)^4 + 12 m_Z^4\right)}{4\pi \sqrt{m_1 m_2} (m_1 + m_2)^2 (m_1 + m_2 - m_h)^2 (m_1 + m_2 + m_h)^2},\\
\langle\sigma v\rangle_{\phi_1 \phi_2\to hh}&\simeq \frac{\lambda_{12}^2 \sqrt{(m_1 + m_2)^2 - 4 m_h^2} \left((m_1 + m_2)^2 + 2 m_h^2\right)^2}{4\pi \sqrt{m_1 m_2} (m_1 + m_2)^2 (m_1 + m_2 - m_h)^2 (m_1 + m_2 + m_h)^2},
\end{aligned}
\end{equation}
and two processes are proportional to $\lambda_{12}^4$:
\begin{equation}
\begin{aligned}
\langle\sigma v\rangle_{\phi_1 \phi_1\to hh}&\simeq \frac{\lambda_{12}^4 v_h^4 \sqrt{m_1^2 - m_h^2}}{\pi m_1^3 \left(m_h^2 - (m_1^2+m_2^2) \right)^2},\\
\langle\sigma v\rangle_{\phi_2 \phi_2\to hh}&\simeq \frac{\lambda_{12}^4 v_h^4 \sqrt{ m_2^2 - m_h^2}}{\pi (\Delta + 1)^3 m_1^3 \left(m_h^2 - (m_1^2+m_2^2)\right)^2}.
\end{aligned}
\end{equation}
The freeze-out temperature is determined by
\begin{equation}
x_f = \ln \frac{0.038 \;g_{\rm eff}\;m_{\rm Pl}\;m_1 \langle\sigma_{\rm eff}v\rangle}{g_*^{1/2}x_f^{1/2}}.
\end{equation}
While the relic abundance is given by
\begin{equation}
\Omega h^2 = \frac{1.07\times 10^9}{g_{*}^{1/2}J(x_f)m_{\rm Pl}({\rm GeV})},
\end{equation}
where $g_*$ is the total degree of freedom of the thermal universe, $m_{\rm Pl}$ is the Planck mass, and the function $J(x_f)$ is defined as
\begin{align}
J(x_f) = \int_{x_f}^{\infty} \frac{\langle \sigma_{\mathrm{eff}} v \rangle}{x^2} \, dx.
\end{align}

We fix the relic density to the observed value $\Omega h^2 = 0.1198$~\cite{ParticleDataGroup:2024cfk} by iteratively determining the freeze-out temperature $x_f$ and the coupling constant $\lambda_{12}$ for various total mass choices $m_{tot} = m_1 + m_2$.
The resulting parameter-space curves in the $(\lambda_{12}, m_{tot})$ plane for $\Delta = 0.05$ (purple line) and $0.07$ (cyan line) are illustrated in the right panel of Fig. \ref{fig:GW-signal} for $m_{tot} \in [60, 200]~\mathrm{GeV}$.
A pronounced dip occurs at $m_{tot} \simeq m_h$ due to the Higgs resonance, while additional sharp features at $m_{tot} \simeq 160~\mathrm{GeV}$ and $m_{tot} \simeq 180~\mathrm{GeV}$ correspond to the opening of the $W^+W^-$ and $ZZ$ production channels, respectively.

\subsection{Existing Constraints}
Apart from the requirement of the DM relic abundance, the interaction between the Higgs boson and the new iDM scalars induces many direct, indirect, and collider search constraints. In this section, we will mainly discuss the limits on the iDM model from DM (in-)direct detection, thermalization of its excited state, monojet+$\slashed{E}_T$ searches, and Higgs precision measurements.
\subsubsection{DM (in-)direct detection}
In our scenario, the DM $\phi_1$ couples to SM only via the $\phi_1\phi_2 H$ interaction, does not directly couple with the SM fermions or gluons, so the tree-level DM-nucleus/electron elastic scattering processes are forbidden. As to the excited state $\phi_2$, after decoupling with the thermal bath in the early universe, it completely decayed into DM and SM particles, leaving no remnants in today's universe. Consequently, no significant signals from $\phi_2$ are expected in direct detection experiments. 
However, the one-loop diagram of two $\lambda_{12}$ vertices can induce an effective $\phi_1\phi_1 h^2$ interaction \cite{Casas:2017jjg}, whose interaction strength is smaller than $\sim (\lambda_{12}/4\pi)^2$. For our interested parameter region, $\lambda_{12}<0.6$, the effective interaction strength should be smaller than $2.3\times10^{-3}$, which is too small to provide any meaningful constraints for direct detection.
Another possibility to consider is whether the upscattering process $\phi_1 N \rightarrow \phi_2 N$ could contribute to direct detection signals. However, the typical mass splitting $\Delta m=m_1 \cdot \Delta \gtrsim 1 \mathrm{GeV}$ is much larger than the kinetic energy of the non-relativistic $\phi_1$, making up-scattering kinematically forbidden. Overall, this iDM model is free from direct detection constraints.

While for the indirect detection, as demonstrated previously, the $\phi_2$ totally decayed before BBN, thus they do not inject energy into the thermal bath during the BBN era. Therefore, they do not change the evolution of the universe. 
While for $m_1>m_h$, DM $\phi_1$ can annihilate to a pair of Higgs bosons via $t$-channel $\phi_2$, which in principle can place a limit on $\lambda_{12}$. However, this process is suppressed by a factor $\lambda_{12}^4\sim 10^{-4}$  \cite{Armand:2022sjf}, thus the cross-section is very small.
For $m_1<m_h$, this process is directly kinematically forbidden. Therefore, due to the non-existent $\phi_2$ in the late universe, the tiny annihilation cross section of $\phi_1\phi_1\to hh$ and the kinematical requirement, the indirect detection cannot impose any restrictions on our interested parameter spaces.

\subsubsection{Monojet+$\slashed{E}_T$ searches}

In our model, both $\phi_1$ and $\phi_2$ couple to the SM via interactions with the Higgs boson, implying potential constraints from collider searches. In particular, the process $p p \rightarrow h^* j \rightarrow \phi_1 \phi_2 j$, followed by $\phi_2 \rightarrow \phi_1 jj$, can be produced at the LHC. However, because the mass difference $\Delta\cdot m_1$ is small, the jets from $\phi_2$ decay are soft and cannot be efficiently reconstructed. Consequently, the final state consists of DM and low-energy jets that escape detection, contributing to a missing transverse energy ($\slashed{E}_T$). Only the initial jet along with the Higgs production is sufficiently energetic to be observed, leading to a mono-jet signature characterized by a high-$p_T$ jet and large $\slashed{E}_T$ ($j + \slashed{E}_T$), distinguishing it from the SM background.

The ATLAS and CMS collaborations have performed extensive searches for DM in the mono-jet plus missing energy channel~\cite{Fox:2011pm,Claude:2022rho,ATLAS:2021kxv,CMS:2021far,CMS:2022qva}. The ATLAS analysis in Ref. \cite{ATLAS:2021kxv}, based on $139~\mathrm{fb}^{-1}$ of data at $\sqrt{s} = 13~\mathrm{TeV}$, requires at least one jet with $p_T(j_1) > 200~\mathrm{GeV}$ and $\slashed{E}_T > 200~\mathrm{GeV}$. At 95\% confidence level (C.L.), the model-independent upper limit on the cross-section for non-SM production is $\sim 0.7~\mathrm{pb}$. \textcolor{black}{While for the CMS \cite{CMS:2021far,CMS:2022qva}, it gives a much weaker constraint than ATLAS due to its smaller data luminosity. Therefore, we will mainly focus on ATLAS.}

By comparing this bound with theoretical predictions obtained using {\tt MadGraph5$\rm \_aMC@NLO$} \cite{Frederix:2018nkq,Alwall:2014hca}, we can constrain the coupling $\lambda_{12}$ for given $m_1$ and $m_2$. In our scenario, for $m_1 + m_2 > m_h$, the corresponding upper limit on $\lambda_{12}$ is larger than the value required to reproduce the observed relic abundance. For example, with $\Delta = 0.1$ and $m_1 = 80~ \mathrm{GeV}$, the predicted cross-section times efficiency computed with {\tt MadGraph} for the relic abundance motivated coupling is $7.6 \times 10^{-5}~\mathrm{pb}$, which is far below the current experimental sensitivity of $0.7~\mathrm{pb}$. Therefore, our model naturally evades existing mono-jet plus missing energy constraints.

\subsubsection{Higgs precision measurements}

Because the scalar fields $\phi_1$ and $\phi_2$ interact with the SM Higgs boson, constraints from Higgs precision measurements must be taken into account. A particularly relevant bound arises from the Higgs total decay width: the combined branching ratio into nonstandard final states is constrained to be below $8.4\%$ at the $95\%$ C.L.~\cite{Cheung:2018ave,Choi:2021nql}. If the mass condition $m_1 + m_2 < m_h$ is satisfied, the SM Higgs can decay directly into $\phi_1$ and $\phi_2$, with the corresponding partial width given by
\begin{equation}\label{eq:decay-width}
    \Gamma(h\to \phi_1\phi_2)=\frac{\lambda_{12}^2 v_h^2 \sqrt{\left((m_h-m_2)^2-m_1^2\right) \left((m_h+m_2)^2-m_1^2\right)}}{16\pi m_h^3}.
\end{equation}
And its branching ratio can be written as
\begin{equation}\label{eq:br-higgs}
    {\rm Br}(h\to \phi_1\phi_2)=\frac{\Gamma(h\to \phi_1\phi_2)}{\Gamma(h\to {\rm SM ~SM})+\Gamma(h\to \phi_1\phi_2)},
\end{equation}
where $\Gamma(h\to {\rm SM ~SM})$ is the Higgs total decay width in SM. By requiring that ${\rm Br}(h\to \phi_1\phi_2)<8.4\%$, the constraint can be obtained, which excludes most of the parameter space where $\lambda_{12} > 5.5 \times 10^{-3}$ and $m_1 + m_2 < m_h$.
If $2m_1 < m_h < m_1 + m_2$, the Higgs can decay via an off-shell $\phi_2$: $h \to \phi_1 \phi_2^* \to \phi_1 \phi_1 + \text{SM SM}$, which is suppressed by the tiny mass splitting and the multi-body phase space. Numerical calculation shows the decay width for $\Delta = 0.1$ is smaller than $10^{-7}~\mathrm{MeV}$ when $\lambda_{12} = 1$, indicating that it cannot impose any meaningful constraint on the parameter space consistent with the DM relic abundance.
If $m_h < 2m_1$, then even the off-shell decay to $\phi_1 \phi_1 + {\rm SM ~SM}$ is kinematically forbidden. Therefore, in summary, for total mass ($m_{tot}=m_1+m_2$) above the Higgs mass, Higgs precision measurements can not provide any effective constraints on the parameter regions corresponding to the correct dark matter relic density.

In addition, the Higgs invisible decay provides a stringent constraint on the coupling $\lambda_{12}$ when $m_1 + m_2 < m_h$. This is because, in this case, the decay products $\phi_1$ and $\phi_2$ typically lead to soft jets or undetectable final states, making them appear as missing energy. The branching ratio for this invisible-like channel has the same form as Eq. (\ref{eq:br-higgs}). Combined with Eq.~(\ref{eq:decay-width}), this invisible decay constraint can be derived.
The projected upper limit on Higgs invisible decay at future CEPC is 0.24\% at the 95\% confidence level \cite{Tan:2020ufz}.
Conservatively, we require $Br(h\to \phi_1\phi_2)<Br(h\to {\rm Inv})<0.24\%$ for the region $m_1 + m_2 < m_h$, which typically requires $\lambda_{12} \lesssim 9 \times 10^{-4}$. These constraints are illustrated in the shaded regions of the right panel in Fig.~\ref{fig:GW-signal}.
In the region where $m_h < 2m_1$, the decay is kinematically forbidden, and therefore, no constraints are applicable in this area. For the mass region, $2m_1<m_h<m_1+m_2$, although the four-body decay process, $h\to \phi_1\phi_2^*\to \phi_1 \phi_1 {\rm SM~ SM}$, is possible, its decay width is too small to impose any constraints due to the heavy suppression from multi-particle final state phase space and heavy mediator mass.

\section{Electroweak Phase Transition and Gravitational Waves}\label{sec:FOPT}

In this section, we examine the vacuum structure and the thermal evolution of the model. For the finite-temperature calculation we use the background fields
\begin{equation}
X=(\varphi_h,\varphi_1,\varphi_2),\qquad
H=\frac{1}{\sqrt{2}}\begin{pmatrix}0\\ \varphi_h\end{pmatrix},
\end{equation}
where $\varphi_{1,2}$ denote the two real dark sector background fields before the mass rotation. The scalar potential at tree level following Eq.~\eqref{eq:Lagrangian-1} is
\begin{equation}
\begin{aligned} 
V_0(X)=&
-\frac{1}{2}\mu^2\varphi_h^2+\frac{\lambda}{4}\varphi_h^4
+\frac{1}{2}\mu_1^2\varphi_1^2+\frac{1}{2}\mu_2^2\varphi_2^2
+\frac{\lambda_\phi}{4}(\varphi_1^2+\varphi_2^2)^2\\
&+\left[\lambda_{I,r}(\varphi_1^2-\varphi_2^2)-2\lambda_{I,i}\varphi_1\varphi_2\right]\varphi_h^2 .
\end{aligned}
\end{equation}
As outlined in Eq. (\ref{eq:parameters}), the model contains four physical parameters $\{ m_1, \Delta, \lambda_\phi, \lambda_{12} \}$. In the relic-density scan, $\lambda_{12}$ is fixed by the observed DM abundance for each $(m_1,\Delta)$, leaving $\lambda_\phi$ as the remaining scalar self-coupling to be scanned. The Lagrangian parameters $\lambda_{I,r}$, $\lambda_{I,i}$, $\mu_1^2$, and $\mu_2^2$ are obtained from the physical inputs through Eq.~(\ref{eq:parameter-relations}).

We impose the vacuum conditions at zero temperature:
\begin{align}
\langle \varphi_1 \rangle =\langle \varphi_2 \rangle= 0, ~~  \langle \varphi_h \rangle = v_h,
\end{align}
ensuring the model reproduces the observed electroweak vacuum. This constraint further restricts the allowed parameter space.

From the physical Lagrangian in Eq.~(\ref{eq:physical-lagrangian}), the off-diagonal Higgs-portal term can become negative along certain field directions and may destabilize the SM vacuum at $(h,\phi_1,\phi_2)=(0,0,0)$. To ensure that electroweak symmetry breaking corresponds to the true vacuum of the theory, we begin our analysis from the physical scalar potential:
\begin{equation}
\begin{aligned}
V(h,\phi_1,\phi_2) = \frac{1}{2}m_1^2\phi_1^2+\frac{1}{2}m_2^2\phi_2^2+\lambda_{12}\phi_1\phi_2(2v_h h+ h^2)+
\lambda_\phi (\phi^\dagger\phi)^2 + V(h).
\end{aligned}
\end{equation}
As an intuitive stability check, we consider the aligned and anti aligned directions, $\phi_2 = \pm\phi_1$. In this limit, the quartic part of the potential simplifies to:
\begin{equation}
\pm\lambda_{12}\phi_1^2 h^2+\lambda_\phi\phi_1^4+\frac{\lambda}{4}h^4 = 
\begin{pmatrix}
h^2&\phi_1^2
\end{pmatrix}
\begin{pmatrix}
\frac{\lambda}{4}&\pm\frac{1}{2}\lambda_{12}\\
\pm\frac{1}{2}\lambda_{12}&\lambda_\phi
\end{pmatrix}
\begin{pmatrix}
h^2\\\phi_1^2
\end{pmatrix}.
\end{equation}
Requiring the determinant of the above coefficient matrix to be positive gives the useful tree-level stability condition along these aligned directions,
\begin{equation}
|\lambda_{12}|<\sqrt{\lambda_\phi\lambda},
\end{equation}
or equivalently $\lambda_\phi>\lambda_{12}^2/\lambda$. In the numerical analysis, this analytic condition is used only as a lower bound for the scan. We further perform a direct zero-temperature minimization of the full zero-temperature effective potential to verify that the electroweak point $(v_h,0,0)$ is the global minimum within the trusted field range.

In the early Universe, the scalar potential receives one-loop zero-temperature and finite-temperature corrections from the hot plasma. We use the full one-loop finite-temperature effective potential with Arnold-Espinosa daisy resummation~\cite{quiros1999finitetemperaturefieldtheory},
\begin{equation}\label{eq:tot-V}
V_{\rm eff}(X,T)
=V_0(X)+V_{\rm CW}(X)+V_{\rm CT}(X)+\Delta V_T(X,T)+V_{\rm daisy}(X,T).
\end{equation}
The Coleman-Weinberg term is
\begin{equation}
V_{\rm CW}(X)=
\sum_\alpha \frac{n_\alpha}{64\pi^2}m_\alpha^4(X)
\left[\ln\frac{|m_\alpha^2(X)|}{Q^2}-c_\alpha\right],
\qquad Q=v_h,
\end{equation}
Here $S_a$ with $a=1,2,3$ denote the three field dependent scalar eigenmodes in the $(\varphi_h,\varphi_1,\varphi_2)$ background space. Their masses are obtained by diagonalizing the scalar Hessian,
\begin{equation}
m_{S_a}^2(X)=\operatorname{Eigenvalues}\left[M_S^2(X)\right],
\qquad
(M_S^2)_{ij}=\frac{\partial^2 V_0}{\partial X_i\partial X_j},
\qquad a=1,2,3,
\end{equation}
and the Coleman-Weinberg sum runs over $\alpha=S_1,S_2,S_3,G,W,Z,t$ with $n_\alpha=(1,1,1,3,6,3,-12)$. The constants are $c_\alpha=3/2$ for scalars and fermions, and $c_\alpha=5/6$ for gauge bosons. The counterterm $V_{\rm CT}$ is chosen such that the one loop zero temperature effective potential preserves the electroweak VEV, the Higgs mass, and the dark sector mass matrix at $(v_h,0,0)$. The Goldstone mass is
\begin{equation}
m_G^2(X)=-\mu^2+\lambda\varphi_h^2
+2\left[\lambda_{I,r}(\varphi_1^2-\varphi_2^2)-2\lambda_{I,i}\varphi_1\varphi_2\right].
\end{equation}
The SM field-dependent masses are $m_W^2(\varphi_h)=m_W^2\varphi_h^2/v_h^2$, $m_Z^2(\varphi_h)=m_Z^2\varphi_h^2/v_h^2$, and $m_t^2(\varphi_h)=m_t^2\varphi_h^2/v_h^2$.

The finite-temperature one-loop term is evaluated with the full thermal functions, not with the high-temperature expansion:
\begin{equation}
\begin{aligned}
\Delta V_T(X,T)=\frac{T^4}{2\pi^2}\Bigg[
&\sum_{a=1}^3J_B\left(\frac{m_{S_a}^2}{T^2}\right)
+3J_B\left(\frac{m_G^2}{T^2}\right)
+6J_B\left(\frac{m_W^2}{T^2}\right)\\
&+3J_B\left(\frac{m_Z^2}{T^2}\right)
-12J_F\left(\frac{m_t^2}{T^2}\right)
\Bigg],
\end{aligned}
\end{equation}
where all masses in $\Delta V_T$ are unresummed field-dependent masses. Daisy resummation is added separately through the bosonic zero modes,
\begin{equation}
\begin{aligned}
V_{\rm daisy}=&-\frac{T}{12\pi}\Bigg\{
\sum_{a=1}^3\left[\mathcal P_{3/2}(\bar m_{S_a}^2)-\mathcal P_{3/2}(m_{S_a}^2)\right]
+3\left[\mathcal P_{3/2}(\bar m_G^2)-\mathcal P_{3/2}(m_G^2)\right]\\
&+2\left[\mathcal P_{3/2}(m_W^2+\Pi_W)-\mathcal P_{3/2}(m_W^2)\right]
+\left[\mathcal P_{3/2}(\bar m_{Z_L}^2)-\mathcal P_{3/2}(m_Z^2)\right]
+\mathcal P_{3/2}(\bar m_{\gamma_L}^2)
\Bigg\},
\end{aligned}
\end{equation}
with $\mathcal P_{3/2}(x)=\operatorname{Re}(x+i0)^{3/2}=x^{3/2}\Theta(x)$. The resummed scalar masses are $\bar m_{S_a}^2=\operatorname{Eigenvalues}[M_S^2+\Pi_S]$ and $\bar m_G^2=m_G^2+\Pi_h$, where
\begin{equation}
\Pi_S=T^2
\begin{pmatrix}
\frac{3g^2+g'^2}{16}+\frac{y_t^2}{4}+\frac{\lambda}{2} & 0 & 0\\
0 & \frac{\lambda_\phi}{3}+\frac{2\lambda_{I,r}}{3} & -\frac{2\lambda_{I,i}}{3}\\
0 & -\frac{2\lambda_{I,i}}{3} & \frac{\lambda_\phi}{3}-\frac{2\lambda_{I,r}}{3}
\end{pmatrix},
\qquad
\Pi_h=T^2\left(\frac{3g^2+g'^2}{16}+\frac{y_t^2}{4}+\frac{\lambda}{2}\right).
\end{equation}
For the longitudinal gauge modes, the charged and neutral sectors are treated separately. The charged modes $W_L^\pm$ do not mix with $B_L$, so their resummed mass is
\begin{equation}
\bar m_{W_L}^2=m_W^2+\Pi_W,\qquad
\Pi_W=\frac{11}{6}g^2T^2 .
\end{equation}
The factor 2 in the gauge Daisy term counts the two charged states $W_L^+$ and $W_L^-$. In the neutral sector, $W_L^3$ mixes with the hypercharge field $B_L$, and the Debye resummed mass matrix in the neutral longitudinal gauge sector is
\begin{equation}
\mathcal M_{L,0}^2=
\begin{pmatrix}
\dfrac{g^2}{4}\varphi_h^2+\Pi_W
&
-\dfrac{gg'}{4}\varphi_h^2
\\
-\dfrac{gg'}{4}\varphi_h^2
&
\dfrac{g'^2}{4}\varphi_h^2+\Pi_B
\end{pmatrix},
\qquad
\Pi_B=\frac{11}{6}g'^2T^2 .
\end{equation}
The two eigenvalues of $\mathcal M_{L,0}^2$ define $\bar m_{Z_L}^2$ and $\bar m_{\gamma_L}^2$. The numerical values of the SM parameters are $m_W = 80.377~{\rm GeV}$, $m_Z = 91.1876~{\rm GeV}$, $m_t = 172.69~{\rm GeV}$, $m_h = 125.25~{\rm GeV}$, and $v_h = 246.22~{\rm GeV}$~\cite{ParticleDataGroup:2024cfk}.

The local minima of $V_{\rm eff}(X,T)$ are traced numerically in the three-dimensional field space as the temperature changes. For each parameter point, all candidate first-order transitions between competing phases are tested for bubble nucleation, and the final zero-temperature phase is required to coincide with the electroweak vacuum $(v_h,0,0)$. The thermal history can therefore be one-step or multi-step, and the transition path may involve both the Higgs and dark sector field directions.

The decay rate per unit volume when phase transition occurs is given by Ref. \cite{quiros1999finitetemperaturefieldtheory}
\begin{equation}
\Gamma(T) \sim T^4 \left( \frac{S_3(T)}{2 \pi T} \right)^{3/2} e^{-S_3(T)/T},
\end{equation}
where $S_3(T)$ denotes the Euclidean action corresponding to the $O(3)$-symmetric bounce solution. A first-order electroweak phase transition takes place when the vacuum decay rate per Hubble volume becomes of order one, signaling the onset of bubble nucleation from the false vacuum. The temperature at which this nucleation begins is defined as the nucleation temperature $T_n$, which satisfies the condition $\Gamma(T_n) = H^3(T_n)$, where $H(T)$ is the Hubble parameter at temperature $T$. 

In a radiation-dominated universe, such as during the electroweak phase transition, the nucleation temperature $T_n$ can be estimated by the following relation~\cite{Quiros:1999jp}
\begin{equation}
\frac{S_3(T_n)}{T_n} \simeq 140,\label{eq:nucleate-temperature}
\end{equation}
which we adopt as the nucleation criterion for a first order phase transition. For each chosen free parameter set, the nucleation temperature $T_n$ is computed by numerically solving Eq.~(\ref{eq:nucleate-temperature}) using the {\tt Python} package {\tt CosmoTransitions}~\cite{Wainwright:2011kj}.

The parameter $\alpha$, which is crucial for determining the strength of the gravitational wave signal, is defined as the ratio of the released vacuum energy to the radiation energy density:
\begin{equation}
\alpha =
\frac{1}{\rho_{\rm rad}(T_n)}
\left[
\Delta V_{\rm eff}(T)-T\frac{d\Delta V_{\rm eff}(T)}{dT}
\right]_{T=T_n},
\qquad
\rho_{\rm rad}(T)=\frac{\pi^2}{30}g_*T^4,
\end{equation}
where $\Delta V_{\rm eff}(T)=V_{\rm eff}(X_{\rm false},T)-V_{\rm eff}(X_{\rm true},T)$ is evaluated between the false and true phases.
Another important parameter, $\beta$, characterizing the inverse time scale of the first order phase transition, is defined as
\begin{equation}
\frac{\beta}{H(T_n)} = T\frac{d\left( S_3(T)/T   \right)}{dT}\Bigg|_{T = T_n}.
\end{equation}

GWs are primarily generated through two processes: sound waves propagating in the plasma and magnetohydrodynamic (MHD) turbulence \cite{Hindmarsh_2015, Hindmarsh_2014}. The total energy density spectrum of the GWs can be expressed as a sum of these two contributions:
\begin{equation}
    \Omega_{GW}h^2\simeq\Omega_{sw}h^2+\Omega_{turb}h^2.
\end{equation}
The component originating from sound waves is given by~\cite{Hindmarsh_2015, Alves:2018jsw}:
\begin{equation}
\Omega_{sw}h^2  = 2.65\times 10^{-6}\left(\frac{H_*}{\beta}\right)\left(\frac{k_v\alpha}{1+\alpha}\right)^2\left(\frac{100}{g_*}\right)^{1/3}\nu_w\left(\frac{f}{f_{sw}}\right)^3\left(\frac{7}{4+3(f/f_{sw})^2}\right)^{7/2},
\end{equation}
where $g_*$ is the relativistic degrees of freedom at the temperature $T_*$, $H_*(T)$ the Hubble constant at the temperature $T_*$ is given by
\begin{align}
H(T_*)=\sqrt{\frac{8\pi G}{3}\times\frac{\pi^2}{30}g_*T_*^4}.
\end{align}
And $\nu_w=1$ is the bubble expansion velocity, 
$k_v= \alpha/\left(0.73+0.083\sqrt{\alpha}+\alpha\right)$
is the fraction of released energy going to the kinetic energy of the plasma~\cite{Espinosa_2010}. Here, the temperature $T_* = T_n$. The peak frequency $f_{sw}$ of the energy density spectrum is
\begin{equation}
f_{sw} = 1.9\times 10^{-5}\frac{1}{\nu_w}\left(\frac{\beta}{H_*}\right)\left(\frac{T_*}{100~\textrm{GeV}}\right)\left(\frac{g_*}{100}\right)^{1/6}\textrm{Hz}.
\end{equation}
In addition to sound waves, a fraction of the energy is released via MHD turbulence. The energy density spectrum for this contribution is ~\cite{Caprini:2009yp, Binetruy:2012ze, Alves:2018jsw}:
\begin{equation}
\Omega_{turb}h^2 = 3.35\times 10^{-4}\left(\frac{H_*}{\beta}\right)\left(\frac{k_{turb}\alpha}{1+\alpha}\right)^{3/2}\left(\frac{100}{g_*}\right)^{1/3}\nu_w\frac{(f/f_{turb})^3}{\left(1+(f/f_{turb})\right)^{11/3}(1+8\pi f /h_*)},
\end{equation}
where the energy proportion of MHD $k_{turb} = 0.1k_v$, and the peak frequency of MHD induced GW is given by
\begin{equation}
f_{turb} = 2.7\times 10^{-5}\frac{1}{\nu_w}\left(\frac{\beta}{H_*}\right)\left(\frac{T_*}{100~\textrm{GeV}}\right)\left(\frac{g_*}{100}\right)^{1/6} \textrm{Hz}.
\end{equation}
And $h_*$ is
\begin{equation}
    h_*=16.5\times10^{-3}\left(\frac{T_n}{100~\text{GeV}}\right)\left(\frac{g_*}{100}\right).
\end{equation}

After getting the energy density spectrum of GWs, the strength of the GW signal detectable by instruments can be quantified using the signal-to-noise ratio (SNR) \cite{Caprini_2016}
\begin{equation}
\textrm{SNR} = \sqrt{\delta\times\mathcal{J}\int df\left[\frac{h^2\Omega_{GW}(f)}{h^2\Omega_{exp}(f)}\right]^2},
\label{eq:SNR}
\end{equation}
where $\mathcal{J}$ represents the mission duration, and $h^2\Omega_{exp}$ characterizes the sensitivity of the detector, here we choose $\mathcal{J}=  9.46 \times 10^7\mathrm{s}$ (more than 3 years) ~\cite{Kawamura:2020pcg}. The factor $\delta$ arises from the number of independent channels available for cross-correlation between detectors, which is equal to 2 in the case of U-DECIGO~\cite{Alves:2018jsw}. 

It should be noted that the parameters adopted in the above GW signal calculations are standard benchmark choices and do not include detailed uncertainty estimations. As discussed in Refs.~\cite{Binetruy:2012ze,Caprini:2009yp,Caprini:2015zlo,Caprini:2019egz,Espinosa:2010hh}, GW spectra from first-order phase transitions are subject to sizable uncertainties, mainly arising from the bubble wall velocity ($v_w$) and the efficiency of vacuum-energy conversion into bulk motion or magnetohydrodynamic (MHD) turbulence ($k_{\mathrm{turb}}$). The wall velocity typically lies in the range $v_w\simeq0.5$–1, depending on the transition strength $\alpha$ \cite{Steinhardt:1981ct}, and can modify the GW amplitude and peak frequency by up to a factor of two. The turbulence efficiency is often parameterized as $k_{\mathrm{turb}}\simeq\epsilon'k_v$ with $\epsilon'\simeq0.05$–0.1 \cite{Caprini:2009yp,Guo:2021qcq}, leading to variations of order unity in $\Omega_{\mathrm{turb}}h^2$. Additional theoretical uncertainties arise from numerical fittings of the sound-wave and turbulence efficiencies, as well as from the modeling of macroscopic thermal parameters and phase-transition dynamics, such as source lifetime and mean bubble separation \cite{Schmitz:2020rag,Hindmarsh:2015qta,Guo:2021qcq,Hindmarsh:2017gnf}. These effects can alter the predicted GW amplitude by up to several orders of magnitude. A comprehensive assessment of these uncertainties has been extensively discussed in the literature; in our study, we therefore adopt benchmark parameter choices to illustrate the characteristic GW features predicted by our model.

\begin{figure}[htb]
\includegraphics[width=0.48\textwidth]{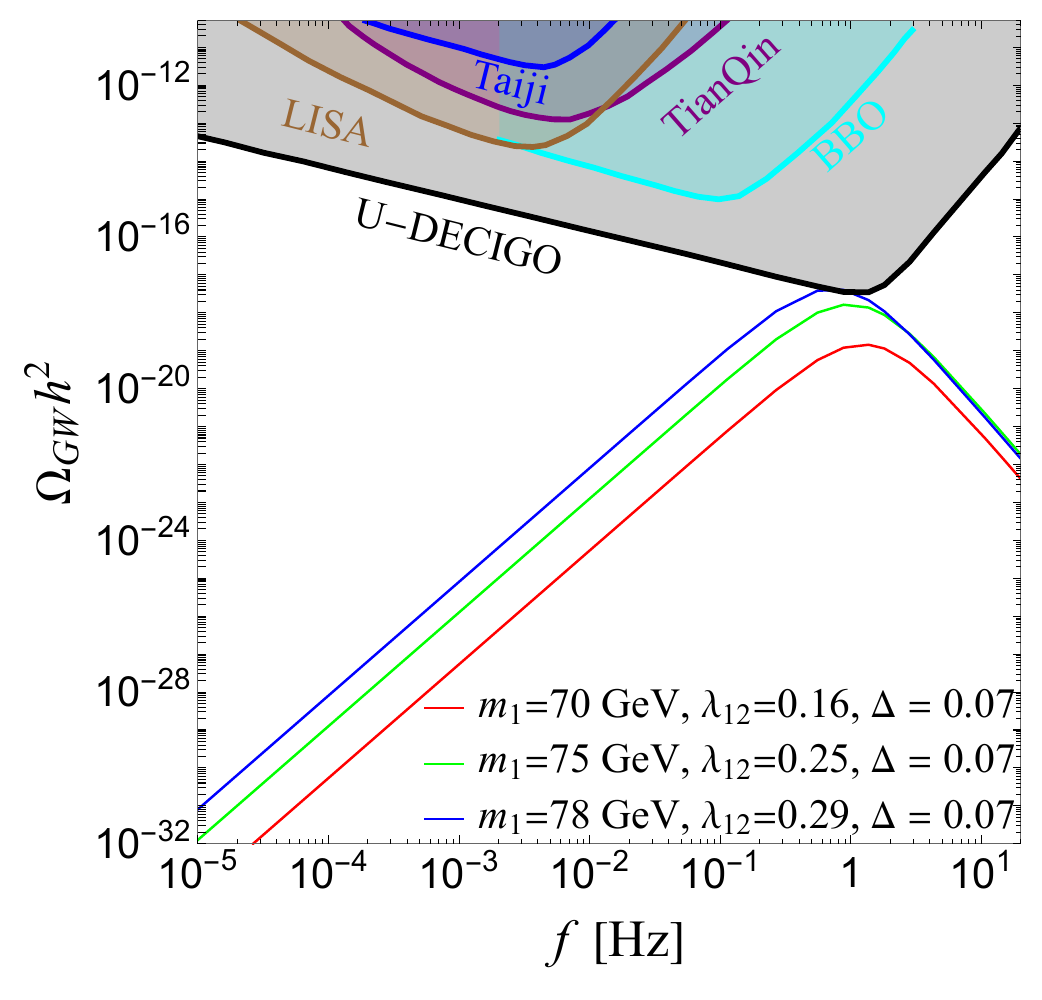}~~~
\includegraphics[width=0.48\textwidth]{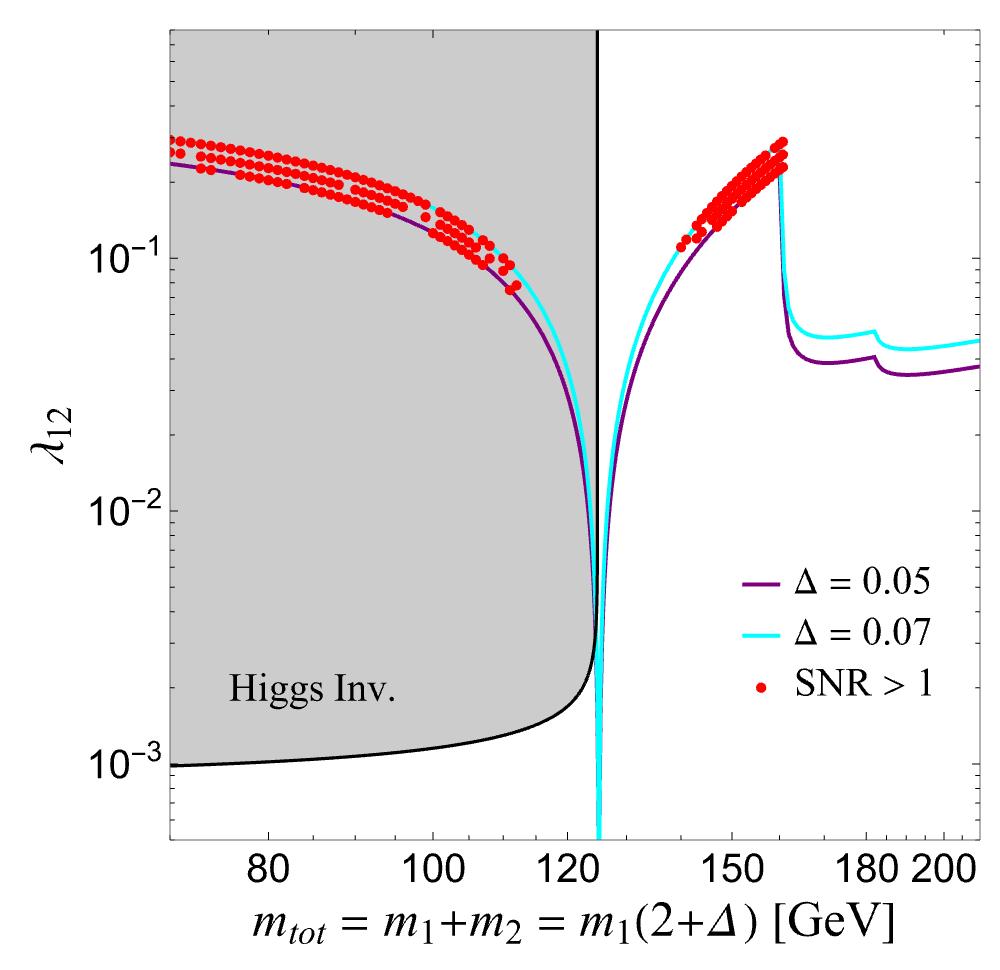}
\caption{
\textbf{Left}: A representative gravitational wave spectrum is shown in comparison with the sensitivity curve of several GW detectors. \\
\textbf{Right}: Red dots represent regions of parameter space that yield strongly first-order phase transitions with a GW signal-to-noise ratio $\mathrm{SNR} > 1$. The purple and cyan curves show the relic abundance targets for $\Delta=0.05$ and $\Delta=0.07$, respectively. The shaded region represents constraints from the Higgs invisible decay.
}
\label{fig:GW-signal}
\end{figure}

We perform the numerical scan using the corrected full finite-temperature effective potential described above. The scan covers $\Delta=0.05,0.06,0.07$ and $m_{\rm tot}=m_1+m_2=60$--$160.75~{\rm GeV}$. For each $(m_{\rm tot},\Delta)$, the coupling $\lambda_{12}$ is fixed by the relic abundance requirement, while $\lambda_\phi$ is scanned near the zero-temperature vacuum-stability boundary. We require the zero-temperature electroweak vacuum before tracing the thermal phases and computing the corresponding GW signal.

\textcolor{black}{The left panel of Fig.~\ref{fig:GW-signal} shows the GW spectra with the sensitivity curves of present and future gravitational wave experiments, including LISA \cite{LISA:2017pwj}, BBO \cite{Harry:2006fi}, Taiji \cite{Hu:2017mde, Luo:2019zal, Luo:2021qji}, TianQin \cite{TianQin:2015yph}, and U-DECIGO. It can be seen that, in some cases, the peak frequency of the gravitational wave energy density generated by the phase transition is close to the frequency where the U-DECIGO detector can reach. As to the other GW detectors, comparing their reaches with our predicted GW spectra, we find that their sensitivity ranges do not overlap with the GW frequencies relevant to our study, and thus are not sufficient to detect the predicted signal strengths. Then, combined with Eq.~(\ref{eq:SNR}), the corresponding $\mathrm{SNR}$ can be calculated. We highlight several representative benchmark points with large $\mathrm{SNR}$ values. Notably, the signals represented by the red and green curves can potentially be detected by both gravitational wave detectors and collider experiments, as will be discussed in the next section.}


The right panel of Fig.~\ref{fig:GW-signal} presents the parameter space leading to strongly first-order phase transitions with $\mathrm{SNR} > 1$, indicated by red dots. It can be observed that these points predominantly originate from the mass range $m_1 \in [40,~80]~\mathrm{GeV}$. The purple and cyan curves represent the relic abundance targets for $\Delta=0.05$ and $\Delta=0.07$, respectively. As implied by Eq.~(\ref{eq:parameter-relations}), a sufficiently large $\lambda_{12}$ can drive $\mu_1^2$ negative, providing a useful diagnostic for a dark field instability. The actual first-order phase transition, however, is determined numerically from the full finite-temperature effective potential. The shaded region denotes parameter space excluded by constraints from Higgs invisible decays. They exclude the mass region for $m_{ tot} < m_h$. 


\section{Inelastic Dark Matter at Colliders}\label{sec:collider}

In our model, two new particles, $\phi_1$ and $\phi_2$, are introduced. We require that $\phi_1$ and $\phi_2$ contribute to the correct relic abundance of DM, support a potential sFOEWPT, and are consistent with existing experimental constraints. These conditions restrict the mass of $\phi_2$ to lie within the range $m_2 \in (m_h - m_1, 2m_W - m_1)$. As discussed in Sec. \ref{sec:int}, $\phi_2$ decays into $\phi_1$ and SM particles, with its lifetime being sufficiently long due to the small mass difference $\Delta$ between $\phi_1$ and $\phi_2$, combined with phase space suppression. This results in $\phi_2$ behaving as a long-lived particle, capable of producing substantial observable signals at $pp$ colliders. Therefore, our focus is on detecting these extended signals from $\phi_2$ within the mass range specified earlier at both ongoing and proposed LHC experiments. 

Specifically, LLP $\phi_2$ may undergo decay subsequent to spatial displacement within the detector, accompanied by a detectable time lag, leading to discernible characteristics distinct from the majority of SM backgrounds. However, to effectively capture and identify these signals, enhancements and specialized apparatus in detectors are imperative. Fortunately, the upcoming HL-LHC upgrade will incorporate precision timing layers that reduce pile-up interference and improve measurements of particle properties, such as position, momentum, and energy. These advancements will be crucial for the exploration of LLPs using tailored detection strategies. Regarding these upgrades, the CMS experiment is developing the Minimum Ionizing Particle (MIP) Timing Detector to achieve these goals~\cite{CERN-LHCC-2017-027,Contardo:2020886}, ATLAS is working on the High Granularity Timing Detector~\cite{Allaire:2018bof}, and LHCb plans to implement similar precision timing upgrades in the near future~\cite{LHCb:2018roe}.

As in many DM scenarios, both $\phi_2$ and $\phi_1$ at colliders typically lead to missing energy, which can be effectively triggered by high-energy monojet events. To efficiently detect LLPs, various strategies are employed. One common approach involves requiring an initial-state radiation (ISR) jet alongside the signal process, which helps identify the primary interaction point \cite{Liu:2018wte}. A high transverse momentum ISR jet with $p_T^j > 120~\mathrm{GeV}$ is typically used to trigger the Jet+MET strategy efficiently \cite{CMS:2014jvv}. Additionally, there have been proposals to include displaced track information in the Level-1 (L1) hardware trigger, with minimum thresholds for track transverse momentum as low as $p_T^j \sim 2~\mathrm{GeV}$ \cite{Bartz:2017nlo}. While using high-energy ISR jets as a trigger mechanism is a cautious approach, there is still room for further refinement. By combining specific signal selection criteria, the behavior of LLPs can be thoroughly explored at the colliders. Moreover, the presence of leptons in the final state can significantly relax the triggering conditions compared to purely hadronic scenarios. To explore LLP $\phi_2$, two promising search methods are adopted: the displaced muon-jet (DMJ) method \cite{Izaguirre:2015zva} and the time-delayed method (TDM) \cite{Liu:2018wte,Berlin:2018jbm}. 

\begin{figure}[htbp]
    \centering
    \includegraphics[width= 0.9 \linewidth,trim = 0 230 80 80,clip]{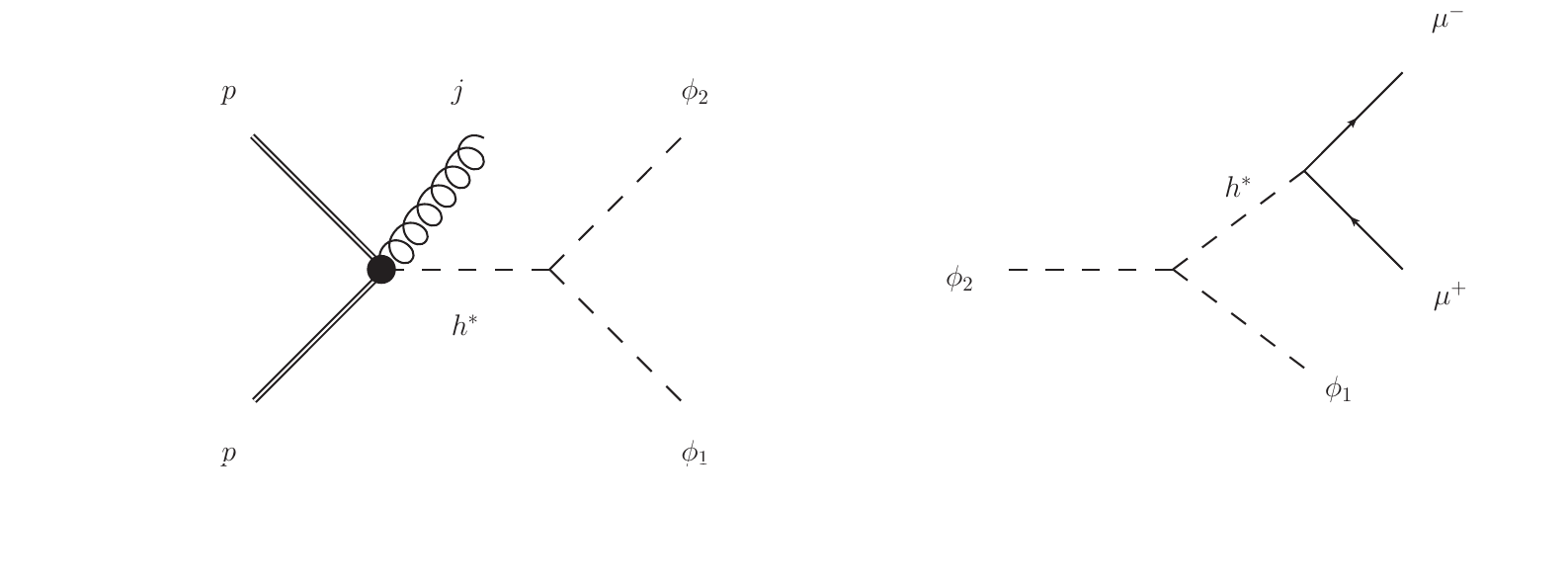}
    \caption{Feynman diagrams for productions and decays of DM partner $\phi_2$ at the HL-LHC.}
    \label{fig:collider-search}
\end{figure}

In our study, we will consider the associated production of $\phi_2$ with an ISR high-$p_T$ jet, followed by the decay of $\phi_2$ to $\phi_1$ and a pair of muons, as shown in Fig. \ref{fig:collider-search}. The full process is described by:
\begin{equation}
    pp\to j h^{*}\to j \phi_2 \phi_1, \phi_2\to\phi_1 \mu^+\mu^-,
\end{equation}
where $h^*$ represents the off-shell Higgs since we focus on the region ($m_1+m_2 > m_h$) relevant for the DM and its partner mass regions of interest. The corresponding Feynman diagrams are shown in Fig. \ref{fig:collider-search}. For simplicity, we focus on the muon decay channel, which, though subdominant, is easily detectable at $pp$ colliders.

\subsection{Displaced Muon-Jet Method}

For the DMJ method, in addition to the stringent ISR jet requirement of $p_T^j > 120~\mathrm{GeV}$, the decay products of $\phi_2$ typically exhibit low momenta due to the small mass difference between $\phi_1$ and $\phi_2$. To identify muons originating from $\phi_2$ decays, we impose that each muon has a transverse momentum $p_T^\mu > 5~\mathrm{GeV}$ \cite{CMS-DP-2014-020}. Furthermore, for accurate reconstruction of their trajectories, these muons must pass through multiple layers of the tracking system, which is ensured by requiring that the radial displacement of the $\phi_2$ decay vertex does not exceed 30 cm. To mitigate the impact of SM backgrounds, a common practice is to require a substantial displacement of muon tracks. Specifically, we impose a transverse impact parameter of $d_0^\mu > 1~\mathrm{mm}$ \cite{Izaguirre:2015zva}. Thus, the comprehensive selection criteria for the DMJ method are:
\begin{equation}
\text{DMJ:} \quad p_T^j > 120~\mathrm{GeV}, \quad p_T^\mu > 5~\mathrm{GeV}, \quad r_{\phi_2} < 30~\mathrm{cm}, \quad d_0^\mu > 1~\mathrm{mm},
\end{equation}
where $r_{\phi_2}$ denotes the radial displacement of the $\phi_2$ decay vertex. As described in Ref. \cite{Izaguirre:2015zva}, these selection criteria are successful in significantly reducing background processes to a negligible extent for an integrated luminosity of HL-LHC, $\mathcal{L}=3{\rm ~ab^{-1}}$.

\subsection{Time-Delayed Method}
For the TDM, the key feature is the slow movement of the long-lived, heavy particle, which leads to a noticeable time delay compared to SM processes. In the SM, particles like mesons and leptons move close to the speed of light, while heavy particles such as gauge bosons and the Higgs boson decay rapidly into lighter particles, causing minimal time delays. However, with heavy new particles beyond the SM, significant time delays occur due to their slower speeds. The time delay, $\Delta t_\mu$, for the decay process $\phi_2 \to \phi_1 \mu^+ \mu^-$ is given by \cite{Liu:2018wte}:
\begin{equation}
\Delta t_\mu = \frac{L_{\phi_2}}{\beta_{\phi_2}} + \frac{L_{\mu}}{\beta_{\mu}} - \frac{L_{\text{SM}}}{\beta_{\text{SM}}},
\end{equation}
where $L$ and $\beta$ represent the propagation distance and velocity of each particle, respectively. For simplicity, we assume the trajectories of $\phi_2$ and its decay products are straight lines, and $\beta_\mu = \beta_{\text{SM}} = 1$. Since particles like $b$ quarks and $\tau$ leptons decay quickly into lighter, highly relativistic particles, these assumptions hold true. The selection criteria for the TDM are:
\begin{equation}
\begin{aligned}
\text{TDM:} \quad & p_T^j > 120~\mathrm{GeV}\; (30~\mathrm{GeV}), \quad p_T^\mu > 3~\mathrm{GeV}, \quad |\eta| < 2.4, \\
& \Delta t_\mu > 0.3~\mathrm{ns}, \quad 5~\mathrm{cm} < r_{\phi_2} < 1.17~\mathrm{m}, \quad z_{\phi_2} < 3.04~\mathrm{m},
\end{aligned}
\end{equation}
where $\eta$ is the pseudorapidity of the jets and muons, and $\Delta t_\mu$ is the measured time delay of the muon. The decay position requirements, $r_{\phi_2}$ and $z_{\phi_2}$, ensure that the decay products leave hits within the CMS MIP Timing Detector. We also evaluate two thresholds for ISR jet transverse momentum: a conservative requirement of $p_T^j > 120~\mathrm{GeV}$ and a more optimistic scenario with $p_T^j > 30~\mathrm{GeV}$, which becomes feasible due to the additional timing and lepton information. Under these conditions, SM background contamination can be neglected.

Using the above two methods, we can explore a dedicated search for the LLP $\phi_2$. The number of signal events arising from $\phi_2$ decays that satisfy the selection criteria is given by:
\begin{equation}
N_{\text{sig}}^{\mu\mu} = \mathcal{L} \cdot \sigma_{\text{sig}} \cdot P(\phi_2) \cdot \epsilon_{\text{cut}},
\end{equation}
where $P(\phi_2)$ is the probability that $\phi_2$ decays while satisfying the specific geometric cuts, $\mathcal{L} = 3~\mathrm{ab}^{-1}$ is the integrated luminosity of the HL-LHC, and $\epsilon_{\text{cut}}$ is the efficiency of the remaining kinematic selection criteria. To accurately determine the signal yield, a parton-level Monte Carlo simulation is performed using {\tt MadGraph5\_aMC@NLO} with the UFO model created by {\tt FeynRules} \cite{Alloul:2013bka}. This simulation samples the decay times of $\phi_2$ for each specific proper lifetime, and the displacement parameters $r_{\phi_2}$ and $d_0^\mu$ (or $\Delta t_\mu$, $r_{\phi_2}$ for DMJ and TDM, respectively) are derived from the kinematics of $\phi_2$ and its decay products. Finally, the efficiency $\epsilon_{\text{cut}}$ is calculated via event-by-event analysis.
\begin{figure}[htbp]
    \centering
    \includegraphics[width=0.48\linewidth]{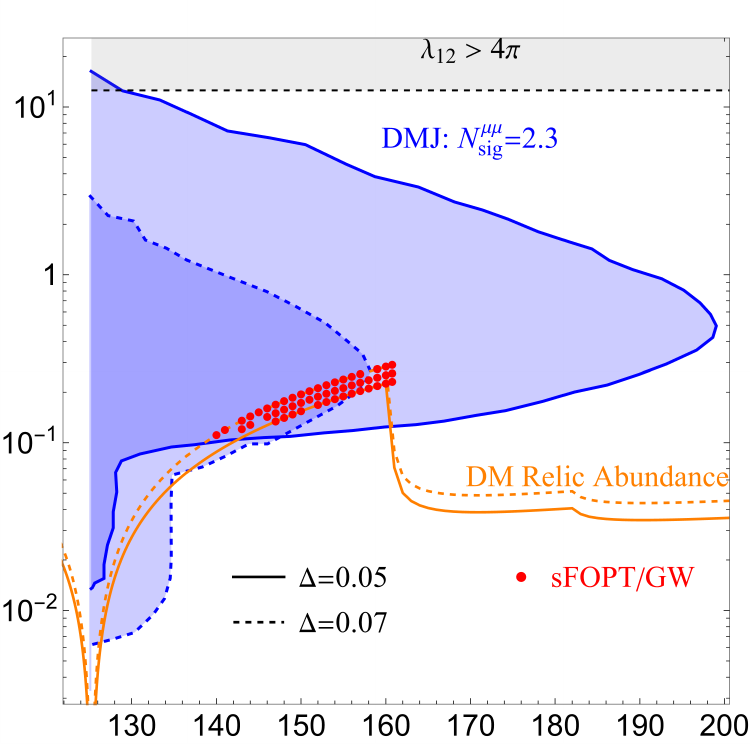}
    \caption{Expected sensitivities at the future HL-LHC to the scalar iDM model in the parameter space determined by $\lambda_{12}$ and $m_{\text{tot}}=m_1+m_2=m_1\cdot(2+\Delta)$, with an integrated luminosity of $\mathcal{L}=3{~\rm ab^{-1}}$ and a center-of-mass energy of $\sqrt{s}=14$ TeV, where the mass splitting $\Delta=0.05$ (dashed) and 0.07 (solid). The projected sensitivity for the displaced muon-jets method with 2.3 signals is depicted by the blue dashed (solid) contour lines. The dashed (solid) orange line and the orange-shaded region depict the parameter space consistent with the observed Dark Matter relic abundance for $\Delta=0.05$ ($\Delta=0.07$) and $\Delta$ ranging from 0.05 to 0.07. The red points labeled ``sFOEWPT/GW'' within this orange-shaded area highlight the regions of strongly FOEWPT with GWs to be explored in forthcoming GW observatories for $\Delta \in [0.05, 0.07]$. Additionally, the gray-shaded region identifies areas where $\lambda_{12}$ surpasses $4\pi$.}
    \label{fig:sensi}
\end{figure}

Based on the analysis of cut efficiency for the DMJ and TDM, we can estimate the sensitivities for the DM partner, $\phi_2$, at HL-LHC, with a center-of-mass energy of $\sqrt{s} = 14$ TeV and an integrated luminosity of $\mathcal{L} = 3~\mathrm{ab^{-1}}$. Unfortunately, the sensitivity of the TDM is too low to probe the relevant parameter space for $\Delta = 0.05$ ($0.07$) in the mass range $m_{tot} \in [m_h, 200]$ ($[m_h, 158]$) GeV and $\lambda_{12} \in [0.014, 17]$ ($[6\times 10^{-3}, 3]$). This limitation arises from the highly boosted nature of $\phi_2$, which results from its small mass and high kinematic energy. Consequently, the time delay required for the TDM method to be effective is not prominent, making this approach less viable for the given parameter space. Additionally, the small mass difference between $\phi_1$ and $\phi_2$ leads to soft muons from $\phi_2$ decays, which complicates the application of the TDM selection criteria.

In contrast, the DMJ method demonstrates significant sensitivity for $\Delta = 0.05$ ($0.07$) in the mass range $m_{tot} \in [m_h, 200]$ ($[m_h, 158]$) GeV and $\lambda_{12} \in [0.014, 17]$ ($[6 \times 10^{-3}, 3]$). As shown in Fig.~\ref{fig:sensi}, the LLP sensitivity of the DMJ method decreases with increasing mass of $\phi_2$. This indicates that the DMJ method is especially effective for relatively lighter DM partners, where the decay products of $\phi_2$ can be more easily detected through distinct kinematic signatures. This decline is due to two key factors: a reduced production cross-section, $\sigma(pp \to j \phi_2 \phi_1)$, and a diminished branching ratio, $Br(\phi_2 \to \phi_1 \mu^+ \mu^-)$, both of which decrease as the mass of $\phi_2$ increases.

Moreover, it is clear that the sensitivity for $\Delta = 0.07$ is considerably weaker than for $\Delta = 0.05$, despite the relatively small difference of 0.02. This disparity is a consequence of the fifth-power dependence of the total decay width of $\phi_2$ on $\Delta$, meaning that even a minor increase in $\Delta$ can lead to a significant increase in the total decay width. As a result, in order to satisfy the long-lived requirement of the DMJ approach, the coupling parameter $\lambda_{12}$ must be substantially reduced. However, this reduction also diminishes the production cross-section of $pp \to j \phi_1 \phi_2$. Therefore, weaker sensitivity is observed for larger mass splitting ratios. Additionally, the branching ratio of $\phi_2 \to \phi_1 \mu^+ \mu^-$ decreases for larger $\Delta$, further impeding the sensitivity of the DMJ method.

Fig.~\ref{fig:sensi} also highlights regions of parameter space that are consistent with the correct DM relic abundance and sFOEWPT. The solid and dashed orange lines represent the boundaries for the correct DM relic abundance for $\Delta = 0.07$ and 0.05, respectively. The red points, labeled ``sFOEWPT/GW'', indicate sFOEWPT regions that are detectable through GWs, which lie within the correct DM relic abundance parameter space for $\Delta \in [0.05, 0.07]$. Furthermore, the gray-shaded area identifies regions where the coupling parameter $\lambda_{12}$ exceeds the theoretical limit of $4\pi$, making the model non-perturbative. Notably, the parameter space in the mass range $m_{tot} \in [142, 155]$ GeV and $\lambda_{12} \in [0.1, 0.3]$ offers a promising opportunity for joint exploration of the iDM model, combining both gravitational wave observations and collider searches.

\section{Conclusions}\label{sec:concl}

The iDM mechanism provides an elegant way to evade stringent direct-detection limits and has attracted growing interest in recent years. While fermionic iDM models have been extensively explored, scalar iDM offers equally rich phenomenology. In this work, we proposed and studied a purely inelastic scalar DM model in which two scalar states-the DM particle $\phi_1$ and its excited partner $\phi_2$-interact inelastically with the SM through a Higgs portal. After diagonalizing the mass matrix, only inelastic couplings among $\phi_1$, $\phi_2$, and the Higgs remain. This setup naturally explains the observed DM relic abundance via thermal coannihilation, predicts a strongly FOEWPT in the early universe capable of generating detectable GWs, and produces long-lived $\phi_2$ signatures at colliders due to the small $\phi_1$–$\phi_2$ mass splitting and the off-shell Higgs mediator.

Crucially, we find out a particularly compelling parameter region, $m_{tot} \in [142, 155]$ GeV and $\lambda_{12} \in [0.1, 0.3]$, where three experimental frontiers converge. In this region, the correct DM relic density is achieved, a sFOEWPT generates GWs detectable by future observatories, and LLP signatures are within the reach of the HL-LHC. This overlap provides a unique opportunity to cross-validate the model through complementary probes from cosmology, astrophysics, and collider physics, highlighting the power of a multi-frontier approach in the search for iDM.

\section{Acknowledgments}
The work of J.G is supported by the Postdoctoral Fellowship Program (Grade C) of China Postdoctoral Science Foundation under Grant No. GZC20252775.
The work of J.L. is supported by the National Science Foundation of China under Grant No. 12235001, No. 12475103 and State Key Laboratory of Nuclear Physics and Technology under Grant No. NPT2025ZX11.
The work of X.P.W. is supported by National Science Foundation of China under Grant No. 12375095, and the Fundamental Research Funds for the Central Universities.
J.L. and X.P.W. thank the Asia Pacific Center for Theoretical Physics (APCTP), Pohang, Korea, for their hospitality during the focus program [APCTP-2025-F01], from which this work greatly benefited. J.L. and X.P.W. also thank the Mainz Institute for Theoretical Physics (MITP) of the PRISMA+ Cluster of Excellence (Project ID 390831469) for its hospitality and partial support during the completion of this work. We also acknowledge with appreciation the valuable discussions and insights provided by the members of the Collaboration of Precision Testing and New Physics.

\bibliographystyle{JHEP}
\bibliography{ref}

\end{document}